\documentclass[12pt,a4paper,oneside]{article}
\usepackage[authoryear]{natbib}
\usepackage{times,epsfig,makeidx,amsfonts,amsmath,dcolumn,enumerate,multirow,graphicx,amssymb,colortbl,bm}
\usepackage[margin=1in]{geometry}

\newtheorem{theorem}{Theorem}

\newtheorem{corollary}{Corollary}

\newtheorem{remark}{Remark}

\newenvironment{proof}[1][Proof]{\noindent\textbf{#1.} }{\ \rule{0.5em}{0.5em}}

\begin{document}
\title{On the Independence Jeffreys prior for skew--symmetric models with applications}

\author{  {\large Francisco Javier Rubio}\footnote{{\sc University of Warwick, Department of Statistics, Coventry, CV4 7AL, UK.} E-mail:  Francisco.Rubio@warwick.ac.uk}  \,{\large and} {\large Brunero Liseo}\footnote{{\sc MEMOTEF, Sapienza Universit{\`a} di Roma.} E-mail: brunero.liseo@uniroma1.it}\\
}
\maketitle

\begin{abstract}
We study the Jeffreys prior of the skewness parameter of a general class of scalar skew--symmetric models. It is shown that this prior is symmetric about $0$, proper, and with tails $O(\lambda^{-3/2})$ under mild regularity conditions. We also calculate the independence Jeffreys prior for the case with unknown location and scale parameters. Sufficient conditions for the existence of the corresponding posterior distribution are investigated for the case when the sampling model belongs to the family of skew--symmetric scale mixtures of normal distributions. The usefulness of these results is illustrated using the skew--logistic model and two applications with real data.
\end{abstract}


\section{Introduction}

The need for modelling data presenting departures from symmetry has fostered the development of distributions that can capture skewness. A popular method to produce this sort of distributions consists of adding a parameter that controls skewness to a symmetric distribution. In this line, \cite{A85} proposed a transformation to produce an asymmetric normal density, termed \emph{skew-normal}, as follows

\begin{eqnarray}\label{skewnormal}
\text{sn}(y;\mu,\sigma,\lambda) = \dfrac{2}{\sigma} \phi\left(\dfrac{y-\mu}{\sigma}\right) \Phi \left(\lambda \dfrac{y-\mu}{\sigma}\right),
\end{eqnarray}

\noindent where $y\in{\mathbb R}$, $\mu\in{\mathbb R}$, $\sigma\in{\mathbb R}_+$, $\lambda\in{\mathbb R}$, $\phi$ is the standard normal probability density function (PDF), and $\Phi$ is the standard normal cumulative distribution function (CDF). The parameter $\lambda$ is often interpreted as a skewness parameter given that the density (\ref{skewnormal}) is asymmetric for $\lambda\neq  0$, and it reduces to the normal PDF for $\lambda=0$. Subsequently, \cite{W04} showed that, in particular, this method can be extended to any continuous symmetric density $f$, with support on ${\mathbb R}$ and mode at $0$, through the transformation

\begin{eqnarray}\label{skewsymmetric}
\text{ss}(y;\mu,\sigma,\lambda) = \dfrac{2}{\sigma} f\left(\dfrac{y-\mu}{\sigma}\right)\pi \left(\lambda \dfrac{y-\mu}{\sigma}\right),
\end{eqnarray}

\noindent where $\pi$, termed \emph{skewing function}, is a function that satisfies $0\leq\pi(y)\leq 1$, and $\pi(-y)=1-\pi(y)$. It follows, then, that any symmetric CDF can be used as a skewing function. Several choices for $f$ and $\pi$ have been explored in the literature, such as the power exponential distribution with power $\delta\in{\mathbb R}_+$ \citep{A86}, the Student-$t$ distribution with $\nu\in{\mathbb R}_+$ degrees of freedom \citep{AC03}, the logistic distribution \citep{N09}, among others. Distributions obtained by means of this method are called \emph{skew--symmetric} distributions. These distributions are widely used nowadays in several contexts such as binary regression \citep{BBB10}, meta--analysis \citep{G12}, data fitting \citep{BGL12}, among many others.

It has been found that several skew--symmetric models present inferential issues. For instance, \cite{A85} showed that the Fisher information matrix of the parameters $(\mu,\sigma,\lambda)$ is singular at $\lambda=0$ for the skew--normal sampling model. In addition, the maximum likelihood estimator of the parameter $\lambda$ can be $\infty$ with positive probability. The cases with infinite estimators are more commonly found in small and moderate samples. These inferential issues are present in other skew--symmetric models \citep{HL12}. Some authors have proposed the use of the Bayesian approach in order to avoid these inferential problems \citep{LL06,BGL12}. In Bayesian practice it is often of interest to employ \emph{noninformative} priors given that they typically produce posterior inference with appealing frequentist properties.  However, due to the singularity of the Fisher information matrix at $\lambda=0$ of some skew--symmetric models, the use of the Jeffreys--rule prior, which is defined as the square root of the determinant of the Fisher information matrix, has been avoided in this kind of models. In addition, the calculation of this sort of prior is typically cumbersome. Reference priors, which are another kind of noninformative priors, have been studied for the skew--normal and the skew Student--$t$ models in \cite{LL06} and \cite{BGL12}. An alternative noninformative prior is the independence Jeffreys prior. This prior is constructed as the product of the Jeffreys priors for each parameter, while treating the remaining parameters as fixed.

In this paper, we study the independence Jeffreys prior associated to the class of skew--symmetric distributions obtained by using a CDF as a skewing function in (\ref{skewsymmetric}). In Section \ref{IJPriors}, we analyse the Jeffreys prior of the skewness parameter $\lambda$ in skew--symmetric models without location and scale parameters. We show that this prior is proper, symmetric about $0$, and with tails $O(\lambda^{-\frac{3}{2}})$ under rather mild regularity conditions. Using these results, we construct the independence Jeffreys prior for the general model with location and scale parameters. In Section \ref{PostExistence} we obtain easy to check sufficient conditions for the propriety of the posterior distribution when the sampling model $f$ in (\ref{skewsymmetric}) belongs to the family of scale mixtures of normal distributions. The case of samples containing censored observations is covered as well. In Section \ref{SkewLogisticSection}, we present the use of these results on the skew--logistic distribution. In Section \ref{Applications}, we illustrate the use of these models in the context of binary regression and stress--strength models. We conclude with some discussion and extensions of this work in Section \ref{DiscussionSection}.


\section{Independence Jeffreys prior for univariate skew--symmetric models}\label{IJPriors}

Throughout we focus on the study of skew--symmetric models of the type

\begin{eqnarray}\label{skewsymmetricInterest}
s(y;\mu,\sigma,\lambda) = \dfrac{2}{\sigma} f\left(\dfrac{y-\mu}{\sigma}\right)G \left(\lambda \dfrac{y-\mu}{\sigma}\right),
\end{eqnarray}

\noindent where $f$ is a continuous symmetric density function with support on ${\mathbb R}$, and $G$ is a CDF with continuous symmetric density $g$ with support on ${\mathbb R}$. This structure covers many cases of practical interest such as the skew--normal distribution \citep{A85}, the skew--$t$ distribution \citep{AC03}, the skew--logistic distribution \citep{N09}, among many others.

Consider first the particular skew--symmetric model (\ref{skewsymmetricInterest}) without location and scale parameters, this is, assuming that $\mu=0$ and $\sigma=1$. Recall that the Fisher information of the parameter $\lambda$ is defined as

\begin{eqnarray*}
I(\lambda)= \int_{\mathbb R} \left[\dfrac{\partial \log s(y;0,1,\lambda)}{\partial \lambda}\right]^2s(y;0,1,\lambda)dy.
\end{eqnarray*}

The Jeffreys prior of the parameter $\lambda$ is defined, up to a proportionality constant, as the square root of the Fisher information $I(\lambda)$, this is, $\pi(\lambda)\propto\sqrt{I(\lambda)}$. The following result characterises the cases where this prior is well--defined at $\lambda=0$.

\begin{remark}\label{FI0}
The Fisher information of $\lambda$ associated to model (\ref{skewsymmetricInterest}), and consequently the Jeffreys prior of $\lambda$, is well--defined at $\lambda=0$ if and only if the second moment of $f$ exists.

\proof See appendix.
\end{remark}

Particular cases of Remark \ref{FI0} have already been reported in the literature. For instance, \cite{BGL12} report the presence of a pole at $\lambda=0$ in the Jeffreys prior of $\lambda$ for the skew Student--$t$ model with $\nu\leq 2$ degrees of freedom. Remark \ref{FI0} shows that this feature is present in many other skew--symmetric models, and that this sort of singularity is linked to the existence of the moments of the underlying symmetric density $f$.

\cite{LL06} and \cite{BGL12} show that the Jeffreys priors of the parameter $\lambda$, for the cases where $f$ and $g$ are normal or Student-$t$ distributions, are proper, decreasing in $\vert \lambda \vert$, and with tails $O(\vert\lambda\vert^{-\frac{3}{2}})$. Their proofs rely upon basic properties of these models, which suggests that there may be other models that lead to a Jeffreys prior of $\lambda$ with the same properties under some reasonable regularity conditions. In order to establish this result, we introduce the following set of sufficient conditions.

\begin{description}
\item[Condition S] Let $f$ and $g$ be continuous density functions with support on ${\mathbb R}$ that satisfy the following conditions:
\begin{enumerate}[(i)]
\item $f$ and $g$ are symmetric about $0$.
\item $f$ is unimodal and there exists a finite constant $M$ such that $0 < f(x) < M$, for all $x\in{\mathbb R}$.
\item The Fisher information $I(\lambda)<\infty$, for all $\lambda\in{\mathbb R}$.
\end{enumerate}
\end{description}

Conditions S.i--S.ii include models of practical interest, such as: the normal distribution, the Student-$t$ distribution, the exponential power distribution, the logistic distribution, among many others. Condition S.iii is simply used to restrict ourselves to those cases where the Jeffreys prior of $\lambda$ exists. Theorem \ref{FI0} provides conditions for the finiteness of $I(0)$, however, for $\lambda\neq 0$ the finiteness of $I(\lambda)$ may require a case by case analysis (for a more detailed study of this point we refer the reader to \citealp{HL12}). The following theorem shows that the results in \cite{LL06} and \cite{BGL12} can be extended to the family of distributions that satisfies Condition S.

\begin{theorem}\label{JPSS}
Let $f$ and $g$ be density functions that satisfy Condition S. Then, the Jeffreys prior of $\lambda$ associated to model (\ref{skewsymmetricInterest}) with $(\mu,\sigma)=(0,1)$ satisfies the following:

\begin{enumerate}[(i)]
\item The Jeffreys prior of $\lambda$ is given by

\begin{eqnarray}\label{JefLambda}
\pi(\lambda) \propto \sqrt{ \int_0^{\infty} x^2 f(x)\dfrac{g(\lambda x)^2}{G(\lambda x)[1-G(\lambda x)]} dx}.
\end{eqnarray}

\item $\pi(\lambda)$ is symmetric about $0$.
\item The tails of $\pi(\lambda)$ are of order $O(\vert\lambda\vert^{-\frac{3}{2}})$.
\item $\pi(\lambda)$ is integrable.
\end{enumerate}
\proof See appendix
\end{theorem}

Based on the tail behaviour, symmetry, and properness of the Jeffreys prior of $\lambda$ shown for the skew--normal model in \cite{LL06}, \cite{BB07} proposed an approximation to this prior using a Student-$t$ distribution with $\nu=1/2$ degrees of freedom and an empirical choice for the scale parameter ($\pi/2$). \cite{BGL12} also proposed a similar approximation for the Jeffreys prior of $\lambda$ of the skew Student-$t$ model. Theorem \ref{JPSS} shows that this approximation might be reasonable in other cases as well. However, the quality of this approximation and the choice for the scale parameter seem to require a case by case analysis. In Section \ref{SkewLogisticSection} we show that this approximation is reasonable for a skew--logistic sampling model.

Condition S.iii can be relaxed to those cases where $I(\lambda)<\infty$ for all $\lambda\neq 0$, possibly leading to an undefined Jeffreys prior at $\lambda=0$ such as those models studied in \cite{BGL12}. The results (ii)--(iii) in Theorem \ref{JPSS} are valid under these relaxed assumptions given that they can be proved using essentially the same technique. The results in \cite{BGL12} also suggest that it is possible to obtain a proper Jeffreys prior $\pi(\lambda)$ for some sampling models despite the singularity of the Fisher information at $\lambda=0$. However, the use of priors containing singularities might be less appealing to practitioners.

We now study the independence Jeffreys prior associated to the skew--symmetric model (\ref{skewsymmetricInterest}) including location and scale parameters. In the next section we also show that this prior leads to a proper posterior distribution under mild conditions.

\begin{theorem}\label{IJPSS}
\item The independence Jeffreys prior of $(\mu,\sigma,\lambda)$ corresponding to a skew--symmetric model (\ref{skewsymmetricInterest}) that satisfies Condition S is given by

\begin{eqnarray}\label{IndJeffreys}
\pi_I(\mu,\sigma,\lambda)\propto \dfrac{1}{\sigma}\pi(\lambda),
\end{eqnarray}

\noindent where $\pi(\lambda)$ is the function defined in (\ref{JefLambda}).

\proof See appendix
\end{theorem}


\section{Existence of the posterior}\label{PostExistence}

In this section, we provide sufficient conditions for the existence of the posterior distribution under the use of the priors studied in the previous section.

\begin{corollary}
Let ${\bf y}=(y_1,\dots,y_n)$ be an \emph{i.i.d.} sample from a skew--symmetric model (\ref{skewsymmetricInterest}) with $(\mu,\sigma)=(0,1)$ that satisfies Condition S. Then, the corresponding posterior distribution of this parameter is proper.

\proof The result follows by the properness of (\ref{JefLambda}) under Condition S.
\end{corollary}

\cite{LL06} show that (\ref{JefLambda}) is proper for the skew--normal sampling model \citep{A85}; and \cite{BGL12} show that this is also the case for the prior (\ref{JefLambda}) associated to a skew--symmetric Student-$t$ sampling model \citep{AC03}. In Section \ref{SkewLogisticSection} we show that the prior (\ref{JefLambda}) associated to a skew--logistic sampling model \citep{N09} is also proper.

For the general model (\ref{skewsymmetricInterest}), with unknown location and scale parameters, the independence Jeffreys prior (\ref{IndJeffreys}) is improper. Then, in order to conduct valid Bayesian inference it is necessary to check conditions for the existence of the corresponding posterior distribution. The following result provides sufficient conditions for the existence of the posterior distribution for the case when $f$ is a scale mixture of normal distributions. The family of scale mixtures of normals contains important models such as the Normal distribution, the Student--$t$ distribution with $\nu$ degrees of freedom, the exponential power distribution with power $1\leq \delta \leq 2$, the logistic distribution, the symmetric $\alpha$-stable family of distributions, among others.

\begin{theorem}\label{PointObs}
Let ${\bf y}=(y_1,\dots,y_n)$ be an \emph{i.i.d.} sample from a skew--symmetric model (\ref{skewsymmetricInterest}) that satisfies Condition S. Suppose also that $f$ is a scale mixture of normals. Then, the posterior distribution of $(\mu,\sigma,\lambda)$ associated to the independence Jeffreys prior (\ref{IndJeffreys}) is proper if $n\geq 2$ and all the observations are different.

\proof See appendix
\end{theorem}

Since the skew--symmetric distributions of interest are continuous, it follows that the probability of obtaining repeated observations is zero. This implies that we can conduct valid Bayesian inference based on this prior whenever $n \geq 2$ for almost any sample. In the Appendix we show that the proof of the propriety of the posterior distribution of $(\mu,\sigma,\lambda)$, under the assumptions in Theorem \ref{PointObs}, can be reduced to proving the propriety of the posterior distribution in the symmetric case. This is, assuming that ${\bf y}$ is an \emph{i.i.d.} sample from a scale mixture of normals $f$ with location and scale parameters $(\mu,\sigma)$ and adopting the prior structure $\pi(\mu,\sigma)\propto \sigma^{-1}$. The propriety of the posterior distribution under the latter assumptions is studied in \cite{FS98}, who also show that the presence of repeated observations in the sample may destroy the existence of the posterior distribution for some scale mixture of normal sampling models. They also present sufficient conditions for the propriety of the posterior distribution in cases when the sample contains repeated observations. We refer the reader to \cite{FS98} for further details on this.

Another scenario of interest is when the observations are recorded as sets of positive probability due to some kind of censoring mechanism. This is, when the collected sample consists of sets $S_1,\dots,S_n$ with ${\mathbb P}(y_i\in S_i)>0$, $i=1,\dots,n$. This framework clearly covers all kinds of interval censoring. The following result shows that the independence Jeffreys prior (\ref{IndJeffreys}) produces a proper posterior distribution in this case as well.

\begin{theorem}\label{CensObs}
Let $S_1,\dots,S_n$ be a sample of censored observations from a skew--symmetric model (\ref{skewsymmetricInterest}) that satisfies Condition S. Suppose also that $f$ is a scale mixture of normals. Then, the posterior distribution of $(\mu,\sigma,\lambda)$ associated to the Bayesian model (\ref{skewsymmetricInterest})--(\ref{IndJeffreys}) is proper if $n\geq 2$ and there exist two sets, say $S_i,S_j$, such that

\begin{eqnarray*}
\inf_{y_i\in S_i,y_j\in S_j}\vert y_i - y_j \vert >0.
\end{eqnarray*}
\proof See appendix
\end{theorem}

This result implies that the posterior distribution of $(\mu,\sigma,\lambda)$ exists whenever the sample of set observations contains at least two observations that do not overlap.

\section{Skew--Logistic model}\label{SkewLogisticSection}

\cite{N09} showed that an interesting member of the skew--symmetric family (\ref{skewsymmetricInterest}) is the skew--logistic distribution, obtained by using the logistic PDF and CDF, $f(t)=\dfrac{e^{-t}}{\left(1+e^{-t}\right)^2}$ and $G(t)=\dfrac{1}{1+e^{-t}}$. The skew--logistic density can be written in closed form, after some algebra, as follows

\begin{eqnarray}\label{SkewLogisticPDF}
\text{sl}(y;\mu,\sigma,\lambda) = \frac{1}{4} \text{sech}^2\left(\frac{y-\mu }{2 \sigma }\right) \left[1+\tanh
   \left(\lambda\dfrac{  y-\mu}{2 \sigma }\right)\right],
\end{eqnarray}

\noindent where $\tanh(\cdot)$ and $\text{sech}(\cdot)$ represent the hyperbolic tangent and the hyperbolic secant functions, respectively. For this sampling model, the Jeffreys prior (\ref{JefLambda}) can be written as indicated below:

\begin{eqnarray}\label{JefLambdaLogistic}
\pi(\lambda) \propto \sqrt{ \int_0^{\infty} x^2 \text{sech}^2\left(\frac{x}{2}\right) \text{sech}^2\left(\frac{\lambda  x}{2}\right) dx}.
\end{eqnarray}

%

\begin{figure}[h]
\begin{center}
\begin{tabular}{c c}
\psfig{figure=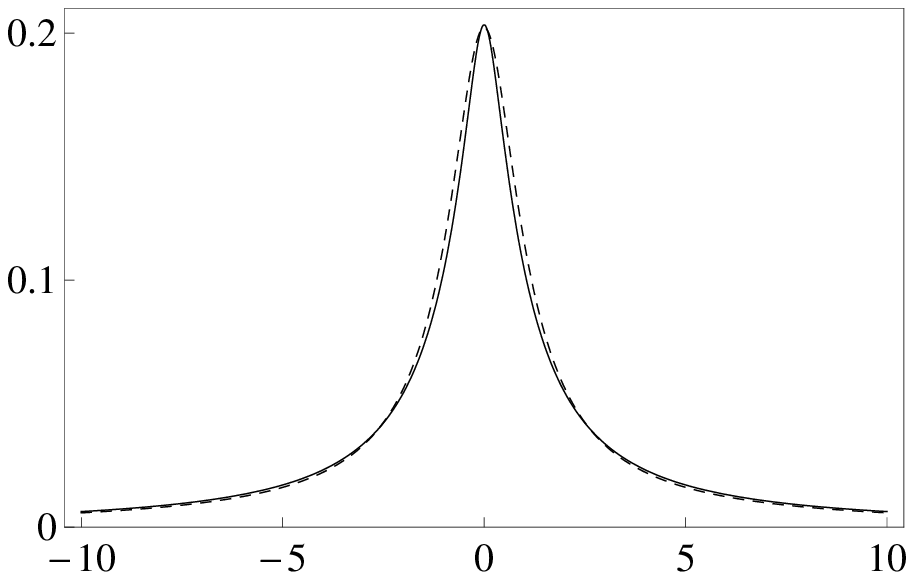,  height=4cm}  &
\psfig{figure=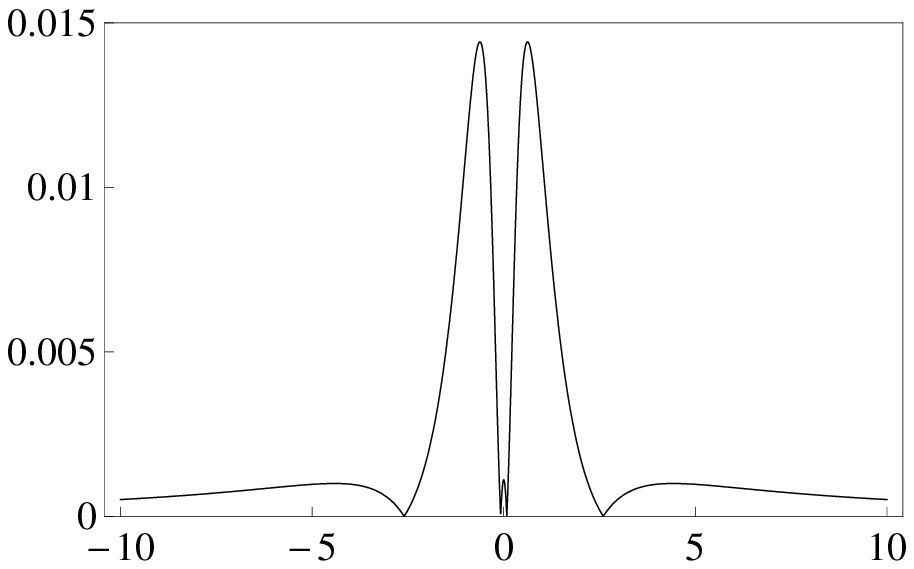,  height=4cm}\\
(a) & (b)
\end{tabular}
\end{center}
\caption{ (a) Jeffreys prior of $\lambda$ (continuous line) and Student-$t$ approximation (dashed line); (b) Absolute difference between the Jeffreys prior of $\lambda$ and the Student-$t$ approximation.}
\label{fig:approx}
\end{figure}

It is easy to check that (\ref{JefLambdaLogistic}) satisfies Condition S and therefore it is proper, as a consequence of Theorem \ref{JPSS}. The tail behaviour, symmetry, and properness shown in this result suggest the use of a Student-$t$ approximation, such as the one proposed in \cite{BB07} for the skew-normal model. Empirically, we have found that $\pi(\lambda)$ can be reasonably well approximated by a Student-$t$ distribution with $1/2$ degrees of freedom and scale parameter $4/3$. Figure $\ref{fig:approx}$ illustrates the quality of this approximation.

For the general skew--logistic model with unknown location and scale parameters it follows that the posterior distribution of $(\mu,\sigma,\lambda)$ using the independence Jeffreys prior (\ref{IndJeffreys}) is proper, given that the logistic distribution can be represented as a scale mixture of normals \citep{S91}, under the conditions in Theorem \ref{PointObs}. Consequently, the results in Theorem \ref{CensObs} also hold for a skew--logistic sampling model.

\subsection{A simulation study}\label{SimulationStudy}

In this section we analyse the empirical coverage of the $95\%$ posterior credible intervals, defined by the 2.5th and 97.5th percentiles, using the independence Jeffreys prior (\ref{IndJeffreys}). We simulate $N= 1000$ data sets of size $n=10,30,100,1000$ from the skew-logistic distribution (\ref{SkewLogisticPDF}) with parameters $\mu_0=0$, $\sigma_0=1$, and $\lambda_0 = 0.5, 1, 2, 5, 10$. The simulation step is implemented by numerical inversion of the corresponding CDF and the probability integral transform. For each of these samples, a posterior sample of size $1000$ was obtained using the $t$-walk algorithm \citep{CF10} with a burn-in period of $10,000$ iterations and a thinning period of $50$ iterations. The proportion of $95\%$ posterior credible intervals that contain the true value of the parameters is reported in Tables \ref{table:n10}--\ref{table:n1000}.
Overall, the frequentist coverage properties of this Bayesian model are good.

\begin{table}[ht]
\begin{center}
\begin{tabular}[h]{|c|c|c|c|c|c|}
\hline
Parameter & $\lambda_0=0.5$ & $\lambda_0=1$ & $\lambda_0=2$ & $\lambda_0=5$ & $\lambda_0=10$\\
\hline
$\mu$ & 0.993 & 0.970 & 0.934 & 0.906 & 0.914\\
\hline
$\sigma$ & 0.933 & 0.965& 0.978 & 0.947 & 0.936 \\
\hline
$\lambda$ & 0.999 & 0.985 & 0.965 & 0.950 & 0.953 \\
\hline
\end{tabular}
\caption{\small Coverage proportions: $n=10$. The value of $\lambda_0$ is specified in each column.}
\label{table:n10}
\end{center}
\end{table}

\begin{table}[ht]
\begin{center}
\begin{tabular}[h]{|c|c|c|c|c|c|}
\hline
Parameter & $\lambda_0=0.5$ & $\lambda_0=1$ & $\lambda_0=2$ & $\lambda_0=5$ & $\lambda_0=10$\\
\hline
$\mu$ & 0.956 & 0.948 & 0.930 & 0.909 & 0.920\\
\hline
$\sigma$ & 0.943 & 0.968 & 0.958 & 0.930 & 0.947 \\
\hline
$\lambda$ & 0.961 & 0.959 & 0.934 & 0.932 & 0.965\\
\hline
\end{tabular}
\caption{\small Coverage proportions: $n=30$. The value of $\lambda_0$ is specified in each column.}
\label{table:n30}
\end{center}
\end{table}

\begin{table}[ht]
\begin{center}
\begin{tabular}[h]{|c|c|c|c|c|c|}
\hline
Parameter & $\lambda_0=0.5$ & $\lambda_0=1$ & $\lambda_0=2$ & $\lambda_0=5$ & $\lambda_0=10$\\
\hline
$\mu$ & 0.965 & 0.952 & 0.937 & 0.920 & 0.934 \\
\hline
$\sigma$ & 0.956 & 0.981 & 0.947 & 0.938 & 0.954 \\
\hline
$\lambda$  & 0.966 & 0.960 & 0.934 & 0.918 & 0.932\\
\hline
\end{tabular}
\caption{\small Coverage proportions: $n=100$. The value of $\lambda_0$ is specified in each column.}
\label{table:n100}
\end{center}
\end{table}

\begin{table}[ht]
\begin{center}
\begin{tabular}[h]{|c|c|c|c|c|c|}
\hline
Parameter & $\lambda_0=0.5$ & $\lambda_0=1$ & $\lambda_0=2$ & $\lambda_0=5$ & $\lambda_0=10$\\
\hline
$\mu$ & 0.958 & 0.946 & 0.943 & 0.947 & 0.946\\
\hline
$\sigma$  & 0.956 & 0.946 & 0.931 & 0.953 & 0.931 \\
\hline
$\lambda$  & 0.952 & 0.950 & 0.938 & 0.944 & 0.949\\
\hline
\end{tabular}
\caption{\small Coverage proportions: $n=1000$. The value of $\lambda_0$ is specified in each column.}
\label{table:n1000}
\end{center}
\end{table}

\newpage
\section{Applications}\label{Applications}

In this section we present two applications of the Bayesian models studied in the previous sections for skew--symmetric distributions using real data. In the first application we consider the use of two skew--symmetric distributions in the context of binary regression to produce a more robust model. Using the Jeffreys prior (\ref{JefLambda}), we propose a hierarchical proper prior based on a modification of the Jeffreys prior proposed in \cite{CIK08} for binary regression models. We show that this prior is scale-invariant with respect to the covariates. The second application consists of a Bayesian model using the prior (\ref{IndJeffreys}) in the context of stress--strength models with dependent observations.

Simulations from the corresponding posterior distributions are obtained using a Metropolis-Hastings algorithm with a burn-in period of $50,000$ iterations and a thinning period of $100$ iterations. Model comparison is conducted using AIC, BIC and Bayes factors. The Bayes factors are calculated using an importance sampling technique. R codes for these examples are available upon request.

\subsection{Binary regression}\label{GenModel}

Binary and binomial observations are common in contexts such as biology, medicine, quality control, among others (see e.g.~\citealp{C99} for a good survey of this). Generalised linear models are a useful tool for modelling this sort of observations given that the probability of observing $y$ successes (failures) of a binomial random variable $Y\in \{0,1,\dots,n\}$ can be related to a certain set of covariates ${\bf x} =(1,x_{1},...,x_{k})^{\top}$, through the model

\begin{eqnarray}\label{glm}
{\mathbb P}\left(Y=y\vert {\bf x}, \bm{\beta}, \bm{\theta} \right) = \binom{n}{y} F\left({\bf x}^{\top}{\bm \beta}; {\bm \theta}\right)^y \left[1-F\left({\bf x}^{\top}{\bm \beta}; {\bm \theta}\right)\right]^{n-y},
\end{eqnarray}

\noindent where $\bm{\beta}=(\beta_0,...,\beta_k)^{\top}$ is a vector of regression coefficients, $F(\cdot;\bm{\theta})$ is a univariate distribution function with shape parameter $\bm{\theta}\in \Theta$, and $F^{-1}$ is called the \emph{link} function. The most common links correspond to the cases where $F$ is the logistic distribution (logit) or the standard normal distribution (probit), which are often referred to as the \emph{canonical links}. It has been found that these links do not always provide a good fit (see e.g.~\citealp{A81}), and also that link misspecification can affect the inference about the parameters \citep{CS92}. An approach for constructing more robust links consists of using a flexible distribution as a link function. In this line we can find a number of references such as \cite{CM77}, \cite{CDS99}, \cite{BM00}, \cite{K08}, \cite{BBB10}, and \cite{WD10}. In this application, we propose a hierarchical prior based on the Jeffreys prior (\ref{JefLambda}) for the generalised linear model (\ref{glm}) with a skew--symmetric link as in (\ref{skewsymmetricInterest}), as described below.

Let $y_i\stackrel{ind.}{\sim}\text{Binomial}\{n_i,S({\bf x}_i^{\top}\bm{\beta};\lambda)\}$, $i=1,\dots,m$, where ${\bf x}_i=(1,x_{i1},...,x_{ik})^{\top}$ is a vector of covariates, $\bm{\beta}=(\beta_0,...,\beta_k)^{\top}$ is a vector of regression coefficients, and $S^{-1}$ is the link function defined by the skew--symmetric distribution (\ref{skewsymmetricInterest}) with $(\mu,\sigma)=(0,1)$ and unknown skewness parameter $\lambda\in{\mathbb R}$. Then, the likelihood function of the parameters $(\bm{\beta},\lambda)$ is given by

\begin{eqnarray}\label{likeglm}
{\mathcal L}(\bm{\beta},\lambda\vert {\bf X},{\bf y}) \propto \prod_{i=1}^m S({\bf x}_i^{\top}\bm{\beta};\lambda)^{y_i}[1-S({\bf x}_i^{\top}\bm{\beta};\lambda)]^{n_i-y_i}.
\end{eqnarray}

\noindent where ${\bf X} = ({\bf x}_1,{\bf x}_2,...,{\bf x}_m)^{\top}$ is the design matrix.

We propose a hierarchical prior structure for the parameters of model $(\ref{likeglm})$, based on a modification of the Jeffreys prior presented in \cite{CIK08}. We adopt the hierarchical prior structure

\begin{eqnarray}\label{priorstructure}
\pi({\bm{\beta},\lambda}) = \pi({\bm{\beta}\vert\lambda})\pi({\lambda}),
\end{eqnarray}

\noindent where

\begin{eqnarray}\label{priorbeta}
\pi(\bm{\beta}\vert\lambda) \propto \det\left[ {\bf X}^{\top} W(\bm{\beta},\lambda) {\bf X}\right]^{\frac{1}{2}},
\end{eqnarray}

\noindent $\det\left[ {\bf X}^{\top} W(\bm{\beta},\lambda) {\bf X}\right]$ denotes the determinant of the matrix ${\bf X}^{\top} W(\bm{\beta},\lambda) {\bf X}$,

\begin{eqnarray*}
W(\bm{\beta},\lambda) &=& \text{diag}(w_1(\bm{\beta},\lambda),...,w_m(\bm{\beta},\lambda)),\\
w_i(\bm{\beta},\lambda) &=& \dfrac{n_i s({\bf x}_i^{\top}\bm{\beta};\lambda)^2}{S({\bf x}_i^{\top}\bm{\beta};\lambda)\{1-S({\bf x}_i^{\top}\bm{\beta};\lambda)\}},
\end{eqnarray*}

\noindent and $\pi({\lambda})$ is given by (\ref{JefLambda}). Since the prior (\ref{priorbeta}) is proper for any $\lambda$ fixed \citep{CIK08}, it follows that the hierarchical structure (\ref{priorstructure}) is also proper if the conditions in Theorem \ref{JPSS} are satisfied. For example, under the use of the skew--normal or the skew--logistic link. In addition, for the skew--normal and the skew--logistic links we can employ the Student-$t$ approximations of $\pi({\lambda})$ described in Section \ref{SkewLogisticSection} in order to facilitate the implementation of prior (\ref{priorstructure}). The proposed prior also presents the following invariance property.

\begin{remark}
Define

\begin{eqnarray*}
C_0({\bf X}) &=& \int_{\mathbb R}\int_{{\mathbb R}^{k+1}} \det\left[ {\bf X}^{\top} W(\bm{\beta},\lambda) {\bf X}\right]^{\frac{1}{2}} \pi(\lambda)d\bm{\beta}d \lambda,\\
C({\bf X}) &=& \int_{\mathbb R}\int_{{\mathbb R}^{k+1}} {\mathcal L}(\bm{\beta},\lambda\vert {\bf X},{\bf y})\det\left[ {\bf X}^{\top} W(\bm{\beta},\lambda) {\bf X}\right]^{\frac{1}{2}} \pi(\lambda)d\bm{\beta}d \lambda,
\end{eqnarray*}

\noindent and ${\bf V}=\operatorname{diag}(1,v_1,...,v_k)$ with $v_1>0,...,v_k>0$. Then, $C_0({\bf X}) = C_0({\bf X}{\bf V})$ and $C({\bf X}) = C({\bf X}{\bf V})$.

\proof The result is a consequence of Theorem 5 from \cite{CIK08}.
\end{remark}

As discussed by \cite{CIK08}, this result implies that $C_0({\bf X})$ and $C({\bf X})$ are scale-invariant with respect to the covariates, which is a desirable property in Bayesian modelling, particularly for conducting variable selection.

\begin{figure}[h]
\centering
\includegraphics[totalheight=2.5in]{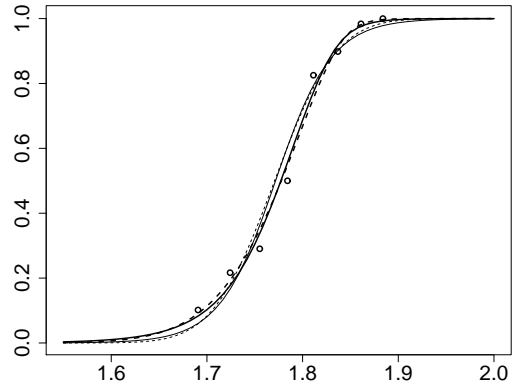}
\caption{ Blist's data. Dose--response curves: logit (solid line), probit (dashed line), skew-logistic link (bold line), skew--normal link (bold dashed line).}
\label{fig:DRC}
\end{figure}

In order to illustrate the use of the proposed Bayesian model, we analyse the popular data set reported in \cite{B35}. The aim of this experiment was to model the response of confused fluor beetles to gaseous carbon disulphide. \cite{A81} mentioned that this data set presents asymmetric departures from the logistic model, hence the use of a skew--symmetric link seems appropriate. Figure \ref{fig:DRC} shows the predictive dose-response curves associated to 4 links: the skew--logistic link with the prior (\ref{priorstructure}), the skew--normal link with the prior (\ref{priorstructure}), the logit link together with the prior $\pi(\bm{\beta}\vert\lambda=0)$ which corresponds to the Jeffreys prior described in \cite{CIK08}, and the probit link together with the prior $\pi(\bm{\beta}\vert\lambda=0)$ which again corresponds to the Jeffreys prior described in \cite{CIK08}. The Bayes factors of the different links against the skew--logistic link, AIC, and BIC values, shown in Table \ref{table:AICBIC}, favour the use of an asymmetric model and slightly favour the skew--logistic link over the other competitors. Table \ref{table:Bliss} shows the predicted observations with these models obtained by multiplying the number of subjects $n_j$ by the predicted probability at the corresponding dose level. This table suggests a better fit of the asymmetric models.

\begin{table}[h]
\begin{center}
\begin{tabular}[h]{|c|c|c|c|c|c|c|c|c|}
\hline
Dose & $n_j$ & $y_{j}$ & logit & probit & skew--logistic & skew--normal\\
\hline
 1.6907 & 59 & 6& 3.5 & 3.5 &  4.9 & 5.4\\
\hline
 1.7242 & 60 &13& 9.9 & 10.8 & 10.6  & 11.6\\
\hline
 1.7552 & 62 &18& 22.5 & 23.5 & 20.9 & 21.6\\
\hline
 1.7842& 56 &28& 33.9 & 33.8 &  31.2 & 30.6 \\
\hline
 1.8113& 63 &52& 50.0 & 49.6 &  48.7  & 47.5 \\
\hline
 1.8369& 59 &53& 53.2 & 53.3 & 54.2 & 53.9 \\
\hline
 1.8610& 62 &61& 59.2 & 59.6 & 60.5 & 60.9 \\
\hline
 1.8839& 60 &60& 58.7 & 59.2 & 59.6 & 59.8\\
\hline
\end{tabular}
\caption{\small Bliss data: predicted observations.}
\label{table:Bliss}
\end{center}
\end{table}

\begin{table}[!h]
\begin{center}
\begin{tabular}[h]{|c|c|c|c|}
\hline
Model & AIC & BIC & Bayes factor\\
\hline
logit & $376.50$ & $376.63$ & 0.12\\
\hline
probit & 375.36 & 375.52 & 0.20\\
\hline
skew-logistic & $370.75$ & $371.00$ & 1\\
\hline
skew-normal & 371.04 & 371.28 & 0.92\\
\hline
\end{tabular}
\caption{\small Bliss data: Model comparison.}
\label{table:AICBIC}
\end{center}
\end{table}

\pagebreak

\subsection{Stress-strength models}

Let $(X,Y)$ be a pair of absolutely continuous random variables such that $Z = X-Y$ has skew--symmetric distribution (\ref{skewsymmetricInterest}). It follows that

\begin{eqnarray}\label{theta}
\theta = {\mathbb P}(X<Y) = {\mathbb P}(X-Y<0) = S(0;\mu,\sigma,\lambda).
\end{eqnarray}

The parameter $\theta$ is called the \emph{stress-strength} coefficient and it has been applied in several contexts (see~\citealp{RS13}). Note that, unlike \cite{RS13} who model the joint distribution of $(X,Y)$, here we are making distributional assumptions on the difference $Z$. In a Bayesian context, if we have a sample of paired observations $(x_i,y_i)$, $i=1,\dots,n$, from $(X,Y)$, then we can obtain a sample from the posterior distribution of $\theta$ by first obtaining a sample from the posterior distribution of $(\mu,\sigma,\lambda)$, obtained in turn by using the sample of differences $z_i=x_i-y_i$, and then by plugging these values into (\ref{theta}). It is worth pointing out that this approach can only be applied when the sample is complete and it does not contain censored observations. For a more general approach that covers these cases see \cite{RS13}.

\begin{figure}[h]
\begin{center}
\begin{tabular}{c c}
\psfig{figure=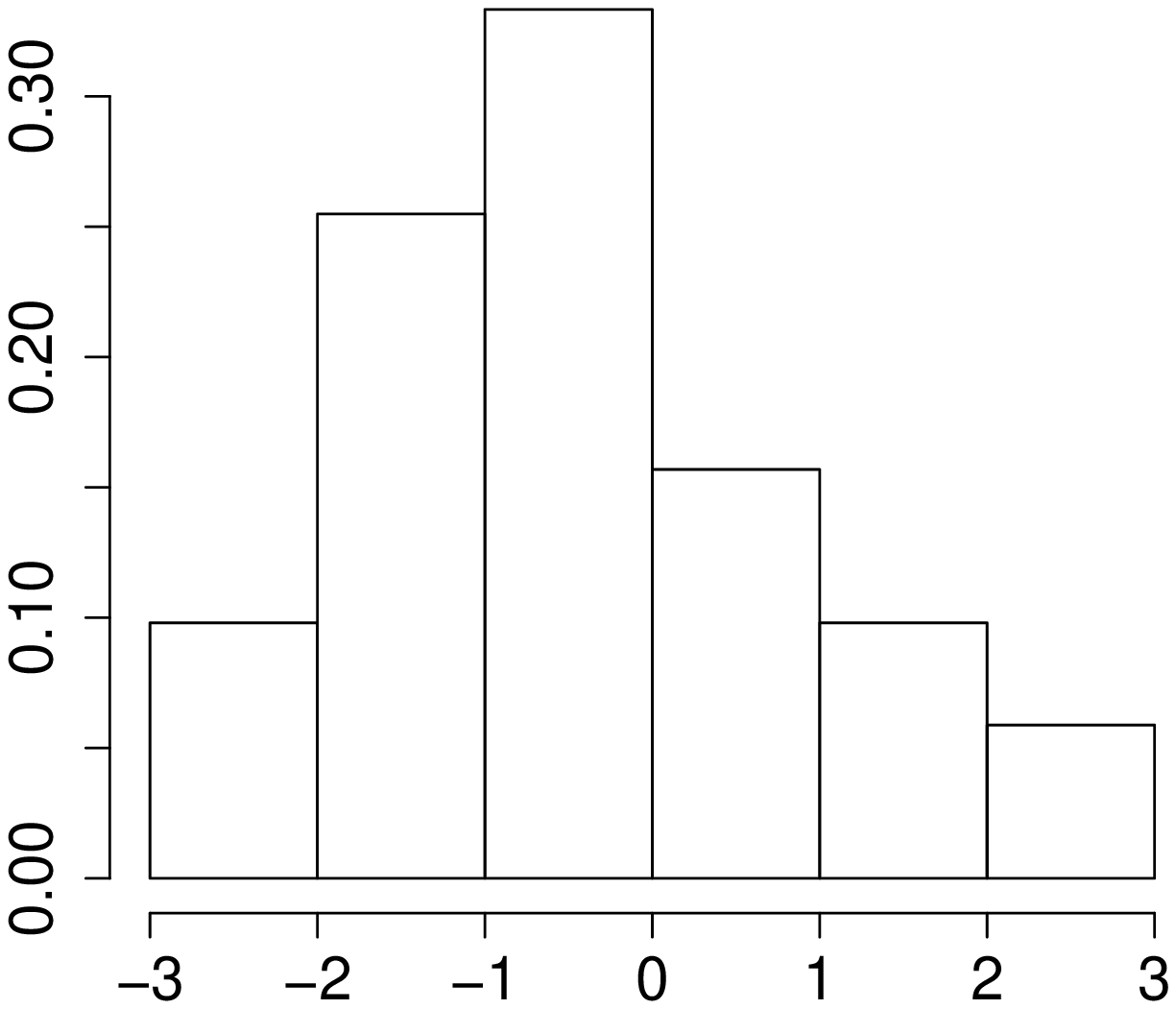,  height=5cm}  &
\psfig{figure=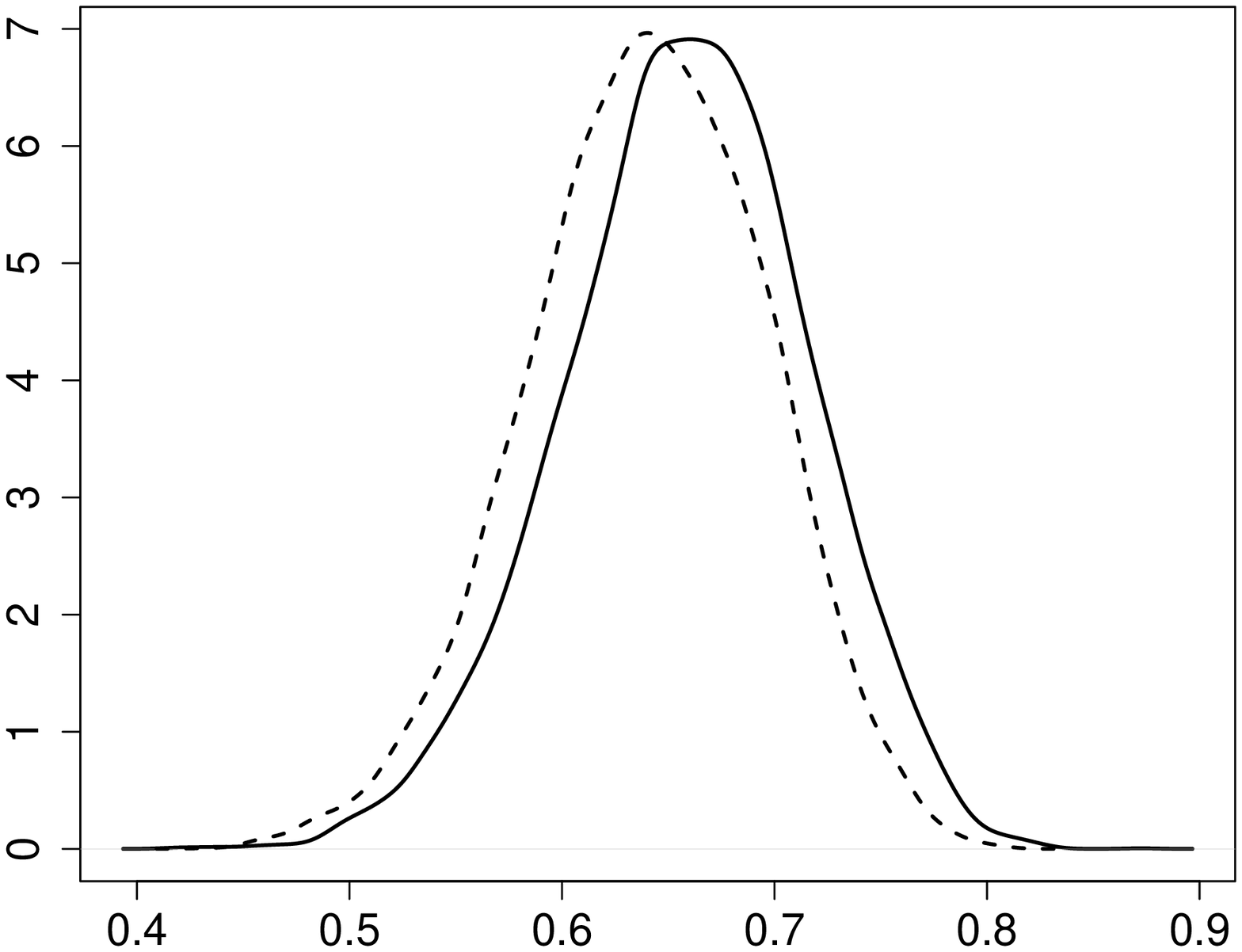,  height=5cm}\\
(a) & (b)
\end{tabular}
\end{center}
\caption{ Melanoma data: (a) Histogram of the difference of the observations; (b) Posterior distribution of $\theta$, skew-logistic model (continuous line), skew--normal model (dashed line).}
\label{fig:Melanoma}
\end{figure}

We consider the data set presented in \cite{V96} which contains 72 lesion scores obtained using both a clinical scheme without a dermoscope ($X$ Test), and a dermoscopic scoring scheme ($Y$ Test). Their main interest was assessing the information provided by the use of the dermoscope. Here, we analyse the subset of $n=51$ non-diseased patients. The sample skewness of the differences $z_i=x_i-y_i$ is $0.57$, and Figure \ref{fig:Melanoma}a shows the histogram of these differences. These features suggest the need for using an asymmetric model. For this purpose, we compare the performance of the skew--normal distribution (\ref{skewnormal}) and the skew-logistic distribution (\ref{SkewLogisticPDF}) together with the independence Jeffreys prior (\ref{IndJeffreys}). Since the sample of differences does not contain repeated observations, it follows that the posterior distribution of $(\mu,\sigma,\lambda)$ is proper and consequently the posterior of $\theta$ is well-defined for both sampling models. Figure \ref{fig:Melanoma}b shows the posterior distributions of $\theta$. We can observe that the inference on $\theta$ under both distributional assumptions is very similar. A $95\%$ posterior credible interval for $\theta$ for the skew--logistic model is $(0.54,0.76)$, while the corresponding interval for the skew--normal model is $(0.52,0.74)$. These intervals do not contain the value $\theta=0.5$, therefore this approach leads to similar results as those obtained in \cite{RS13}. The Bayes factors of the models of interest against the skew--logistic model, AIC, and BIC values, shown in Table \ref{table:AICBICSS}, slightly favour the skew--normal model.

\begin{table}[!h]
\begin{center}
\begin{tabular}[h]{|c|c|c|c|}
\hline
Model & AIC & BIC & Bayes factor\\
\hline
skew-logistic & 176.17 & 181.96 & 1\\
\hline
skew-normal & 175.01 & 180.80 & 1.5\\
\hline
\end{tabular}
\caption{\small Melanoma data: Model comparison.}
\label{table:AICBICSS}
\end{center}
\end{table}

\pagebreak
\section{Discussion}\label{DiscussionSection}

We have studied the Jeffreys prior of the skewness parameter of a general class of scalar skew--symmetric models as well as the independence Jeffreys prior for the same class of models with unknown location and scale parameters. We have shown that this sort of priors has appealing properties such as symmetry, properness, and identifiable tail behaviour that allow in many cases a tractable approximation that facilitate their implementation. We have also presented easy to check conditions for the existence of the posterior distribution for a general subclass of skew--symmetric sampling models. Given that the prior on the skewness parameter has heavy tails, $O(\vert\lambda\vert^{-3/2})$, it is expected to obtain good frequentist properties of the corresponding Bayesian models since heavy--tailed priors are usually employed as ``vague priors''. This feature was illustrated using a simulation study in Section \ref{SimulationStudy}.

One of the unpleasant properties of the priors studied in this paper is that they are well--defined at $\lambda=0$ only when the transformed symmetric density $f$ has a finite second moment. This can be considered as a limitation for the use of these Bayesian models. However, things must be considered in perspective. In some cases where our prior cannot be defined at zero, maximum likelihood estimation fails also, and we do not know any other broadly satisfactory alternative method. Inspired by the structure of the independence Jeffreys prior (\ref{IndJeffreys}) we can construct a more general benchmark prior for skew--symmetric models as follows

\begin{eqnarray}\label{benchmark}
\pi(\mu,\sigma,\lambda)\propto \sigma^{-1}p(\lambda),
\end{eqnarray}

\noindent where $p(\lambda)$ is any proper prior. Using this prior structure, the corresponding posterior distribution is proper for any skewing function $G$ if $f$ in (\ref{skewsymmetricInterest}) is a scale mixture of normals, the sample size $n\geq 2$, and all the observations are different. The proof of this result is similar to that of Theorem \ref{PointObs}. The study of appropriate choices for $p(\lambda)$ is a matter of further research.

Extensions to this work include the study of the independence Jeffreys prior of multivariate skew--symmetric models as well as the propriety of the corresponding posterior distributions.  Also, the technique employed in the proof of the propriety of the posterior distribution in Theorem \ref{PointObs} can be extended, with some care, to the use of the prior structures (\ref{IndJeffreys}) and (\ref{benchmark}) in the context of linear regression models with skew--symmetric residual errors using the results in \cite{FS00}. This opens the door to a variety of applications.

\pagebreak

\section*{Appendix}

\subsection*{Proof of Remark \ref{FI0}}

First, after some algebra we get

\begin{eqnarray*}
\dfrac{\partial \log[s(x;0,1,\lambda)]}{\partial \lambda} = \dfrac{x g(\lambda x)}{G(\lambda x)}.
\end{eqnarray*}

Using this calculation it follows that the Fisher information of $\lambda$ can be written as follows

\begin{eqnarray*}
I(\lambda) = \int_{-\infty}^{\infty} \left[\dfrac{x g(\lambda x)}{G(\lambda x)}\right]^2 2f(x)G(\lambda x) dx = 2\int_{-\infty}^\infty x^2 f(x) \dfrac{g(\lambda x)^2}{G(\lambda x)}dx.
\end{eqnarray*}

Therefore $I(0)=4g(0)^2\int_{\mathbb R} x^2 f(x)dx$, which establishes the relationship of $I(0)$ and the finiteness of the second moment of $f$.

\subsection*{Proof of Theorem \ref{JPSS}}
\begin{enumerate}[(i)]

\item  Splitting the integration range in the expression of $I(\lambda)$, using the symmetry of $f$ and $g$, and the equality $G(-x)=1-G(x)$ we obtain

\begin{eqnarray}\label{FisherInfo}
I(\lambda) &=& 2\int_{-\infty}^0 x^2 f(x) \dfrac{g(\lambda x)^2}{G(\lambda x)}dx + 2\int_0^{\infty} x^2 f(x) \dfrac{g(\lambda x)^2}{G(\lambda x)}dx \notag\\
&=& 2\int_0^{\infty} x^2 f(x) \dfrac{g(\lambda x)^2}{G(\lambda x)[1-G(\lambda x)]}dx.
\end{eqnarray}

The result follows by taking the square root of the latter expression.

\item The symmetry of $\pi(\lambda)$ follows from that of $I(\lambda)$, which in turns is a consequence of the symmetry of $f$ and $g$, and the equality $G(-x)=1-G(x)$.

\item  Using the fact that $f$ is upper bounded we can obtain the following upper bound for the Fisher information of $\lambda$

$$I(\lambda) = 2\int_0^{\infty} x^2 f(x) \dfrac{g(\lambda x)^2}{G(\lambda x)[1-G(\lambda x)]}dx \leq 2M \int_0^{\infty} x^2  \dfrac{g(\lambda x)^2}{G(\lambda x)[1-G(\lambda x)]}dx.$$

Now, consider the change of variable $u= \lambda x$, with $\lambda>0$, then we can rewrite this upper bound as follows

\begin{eqnarray}\label{utails}
I(\lambda) \leq \dfrac{2M}{\lambda^3} \int_0^\infty u^2\dfrac{g(u)^2}{G(u)[1-G(u)]}du,
\end{eqnarray}

After the same change of variable $u=\lambda x$ we can rewrite the Fisher information as

\begin{eqnarray*}
I(\lambda) = \dfrac{2}{\lambda^3}\int_0^{\infty} u^2f\left(\dfrac{u}{\lambda}\right) \dfrac{g^2(u)}{G(u)[1-G(u)]} \,du.
\end{eqnarray*}

Note that for $\lambda\geq L>0$, $f\left(\dfrac{u}{\lambda}\right) \geq f\left(\dfrac{u}{L}\right)$, for all $u>0$. Then

\begin{eqnarray}\label{ltails}
I(\lambda) \geq \dfrac{2}{\lambda^3}\int_0^{\infty} u^2 f\left(\dfrac{u}{L}\right)\dfrac{g^2(u)}{G(u)[1-G(u)]} \,du,
\end{eqnarray}

\noindent for $\lambda>L>0$. Therefore, by combining $(\ref{utails})$ and $(\ref{ltails})$ it follows that $I(\lambda)$ has tails of order $O(\vert\lambda\vert^{-3})$ which implies that $\pi(\lambda)$ has tails $O(\vert\lambda\vert^{-3/2})$.

\item Let $a>0$, using that $\pi(0)<\infty$ (as a consequence of Condition S.iii), and that $\pi(\lambda)$ is finite on $[0,\infty)$ by assumption, it follows that

\begin{eqnarray}\label{Int1}
\int_0^a\pi(\lambda) d\lambda <\infty.
\end{eqnarray}

Now, using $(\ref{utails})$ we have that

\begin{eqnarray}\label{Int2}
\int_a^{\infty}\pi(\lambda) d\lambda \leq C\int_a^{\infty} \dfrac{ 1}{\lambda^{\frac{3}{2}}} d\lambda <\infty,
\end{eqnarray}

\noindent where $C$ is a positive constant. Combining (\ref{Int1}) and (\ref{Int2}), and using that $\pi(\lambda)$ is symmetric about $0$, the properness of $\pi(\lambda)$ follows.

\end{enumerate}

\subsection*{Proof of Theorem \ref{IJPSS}}

The diagonal entries of the Fisher information matrix of $(\mu,\sigma,\lambda)$ are given by

\begin{eqnarray*}
I_{\mu,\mu} &=& \dfrac{2}{\sigma^2}\int_{-\infty}^{\infty} \left[\dfrac{f^{\prime}(t)}{f(t)}+\lambda\dfrac{g(\lambda t)}{G(\lambda t)}\right]^2f(t)G(\lambda t) dt,\\
I_{\sigma,\sigma} &=& \dfrac{2}{\sigma^2}\int_{-\infty}^{\infty} \left[1 + t\dfrac{f^{\prime}(t)}{f(t)}+\lambda t \dfrac{g(\lambda t)}{G(\lambda t)}\right]^2f(t)G(\lambda t) dt,\\
I_{\lambda,\lambda} &=& \int_{-\infty}^{\infty} t^2f(t)\dfrac{g(\lambda t)^2}{G(\lambda t)} dt.
\end{eqnarray*}

Note that $I_{\mu,\mu}$ does not depend on $\mu$; $I_{\sigma,\sigma}$ depends on $\sigma$ through the factor $\sigma^{-2}$; and $I_{\lambda,\lambda}$ coincides with $I(\lambda)$, defined in the proof of Theorem \ref{JPSS}. The result follows from these observations and by the definition of the independence Jeffreys prior.

\subsection*{Proof of Theorem \ref{PointObs}}

First of all, recall that a posterior distribution is proper whenever the marginal distribution $p(y_1,\dots,y_n)<\infty$ \citep{FS98}. Now, note that

\begin{eqnarray}\label{ineqss}
s(y;\mu,\sigma,\lambda) \leq \dfrac{2}{\sigma}f\left(\dfrac{y-\mu}{\sigma}\right).
\end{eqnarray}

Then, if follows that

\begin{eqnarray*}
p(y_1,\dots,y_n) &=& \int_{\mathbb R} \int_{{\mathbb R}_+} \int_{\mathbb R} \prod_{j=1}^n \left[ s(y_j;\mu,\sigma,\lambda)\right] \dfrac{\pi(\lambda)}{\sigma}d\mu d\sigma d\lambda\\
&\leq& \int_{{\mathbb R}_+} \int_{\mathbb R} \left[\prod_{j=1}^n \dfrac{2}{\sigma}f\left(\dfrac{y_j-\mu}{\sigma}\right)\right]\dfrac{1}{\sigma}d\mu d\sigma \int_{\mathbb R} \pi(\lambda)d\lambda
\end{eqnarray*}

By Theorem \ref{JPSS}, we have that $\pi(\lambda)$ is proper. Then, it follows that the posterior distribution of $(\mu,\sigma,\lambda)$ exists whenever the posterior distribution of $(\mu,\sigma)$ exists for a scale mixture of normals sampling model and the prior $\pi(\mu,\sigma)\propto \sigma^{-1}$. The properness of the latter, for $n\geq 2$, follows by Theorem 1 from \cite{FS98}.

%
%
%
%
%

\subsection*{Proof of Theorem \ref{CensObs}}

The result follows again by using inequality (\ref{ineqss}) and Theorem 4 from \cite{FS98}.



\begin{thebibliography}{9}
\bibitem[Aranda-Ordaz(1981)]{A81} Aranda-Ordaz, F. J. (1981). On two families of transformations to additivity for binary response data. {\sl Biometrika} 68: 357--363.
\bibitem[Azzalini(1985)]{A85} Azzalini, A. (1985). A class of distributions which includes the normal ones. {\sl Scandinavian Journal of Statistics} 12: 171–-178.
\bibitem[Azzalini(1986)]{A86} Azzalini, A. (1986). Further results on a class of distributions which includes the normal ones. {\sl Statistica} 46: 199--208.
\bibitem[Azzalini and Capitanio(2003)]{AC03} Azzalini, A. and Capitanio, A. (2003). Distributions generated by perturbation of symmetry with emphasis on a multivariate skew-t distribution. {\sl Journal of Royal Statistical Society}, Series B 65: 367--389.
\bibitem[Bayes and Branco(2007)]{BB07} Bayes, C. L. and Branco, M. D. (2007). Bayesian inference for the skewness parameter of the scalar skew-normal distribution. {\sl Brazilian Journal of Probability and Statistics} 21: 141--163.
\bibitem[Basu and Mukhopadhyay(2000)]{BM00} Basu, S. and Mukhopadhyay, S. (2000). Bayesian analysis of binary regression using symmetric and asymmetric links. {\sl Sankhya: The Indian Journal of Statistics}, Series B 62: 372--387.
\bibitem[Baz{\'a}n et al.(2010)]{BBB10} Baz{\'a}n, J. L., Bolfarine, H. and Branco, M. (2010). A framework for skew-probit links in binary regression. {\sl Communications in Statistics -- Theory and Methods} 39: 678--697.
\bibitem[Bliss(1935)]{B35} Bliss, C. I. (1935). The calculation of the dosage-mortality curve. {\sl Annals of Applied Biology} 22: 134-–167.
\bibitem[Branco et al.(2012)]{BGL12} Branco, M. D., Genton, M. G. and Liseo, B. (2012). Objective Bayesian Analysis of Skew-t
Distributions. {\sl Scandinavian Journal of Statistics} 40: 63–-85.
\bibitem[Chen et al.(1999)]{CDS99} Chen, M. H., Dey, D. K. and Shao, Q. M. (1999). A new skewed link model for dichotomous quantal response data. {\sl Journal of the American Statistical Association} 94: 1172--1186.
\bibitem[Chen et al.(2008)]{CIK08} Chen, M. H., Ibrahim, J. G. and Kim, S. (2008). Properties and implementation of Jeffreys's priors in binomial regresion models. {\sl Journal of the American Statistical Association} 103: 1659--1664.
\bibitem[Christen and Fox(2010)]{CF10} Christen, J. A. and Fox, C. (2010). A general purpose sampling algorithm for continuous distributions (the t-walk). {\sl Bayesian Analysis} 5: 263--282.
\bibitem[Collet(1999)]{C99} Collett, D. (1999). {\sl Modelling Binary Data}. Chapman \& Hall/CRC, Boca Raton Florida.
\bibitem[Copenhaver and Mielke(1977)]{CM77} Copenhaver, T. W. and Mielke, P. W. (1977). Quantit analysis: a quantal assay refinement. {\sl Biometrics} 33: 175--186.
\bibitem[Czado and Santner(1992)]{CS92} Czado, C. and Santner, T. J. (1992). The effect of link misspecification on binary regression inference. {\sl Journal of Statistical Planning and Inference} 33: 213--231.
\bibitem[Fern\'andez and Steel(1998)]{FS98} Fern{\'a}ndez, C. and Steel, M. F. J. (1998). On the dangers of modelling through continuous distributions: A Bayesian perspective, in Bernardo, J. M., Berger, J. O., Dawid, A. P. and Smith, A. F. M. eds., {\sl Bayesian Statistics 6}, Oxford University Press (with discussion), pp.~213--238.
\bibitem[Fern\'andez and Steel(2000)]{FS00} Fern{\'a}ndez, C. and Steel, M. F. J. (2000). Bayesian Regression Analysis with Scale Mixtures of Normals. {\sl Econometric Theory} 16: 80--101.
\bibitem[Guolo(2012)]{G12} Guolo, A. (2012). Flexibly modeling the baseline risk in meta-analysis. {\sl Statistics in Medicine}, in press.
\bibitem[Hallin and Ley(2012)]{HL12} Hallin, M. and Ley, C. (2012). Skew-symmetric distributions and Fisher information --– a tale of two densities. {\sl Bernoulli} 18: 747--763.
\bibitem[Kim et al.(2008)]{K08} Kim, S., Chen, M. H. and Dey, D. K. (2008). Flexible generalized $t-$link models for binary response data. {\sl Biometrika} 95: 93--106.
\bibitem[Liseo and Loperfido(2006)]{LL06} Liseo, B. and Loperfido, N. (2006). A note on reference priors for the scalar skew-normal distribution. {\sl Journal of Statistical  Planning and Inference} 136: 373--389.
\bibitem[Nadarajah(2009)]{N09} Nadarajah, S. (2009). The skew logistic distribution. {\sl Advances in Statistical Analysis} 93: 197-–203.
\bibitem[Rubio and Steel(2013)]{RS13} Rubio, F. J. and Steel, M. F. J. (2013). Bayesian Inference for $P(X<Y)$ using Asymmetric Dependent Distributions. {\sl Bayesian Analysis} 8: 43–-62.
\bibitem[Stefanski(1991)]{S91} Stefanski, L. A. (1991). A normal scale mixture representation of the logistic distribution. {\sl Statistics \& Probability Letters} 11: 69--70.
\bibitem[Venkatraman and Begg(1996)]{V96} Venkatraman, E. S. and Begg, C. B. (1996). A distribution--free procedure for comparing operating characteristic curves from a paired experiment. {\sl Biometrika} 83: 835--848.
\bibitem[Wang et al.(2004)]{W04} Wang, J., Boyer, J. and Genton M. C. (2004). A skew symmetric representation of multivariate distributions. {\sl Statistica Sinica} 14: 1259--1270.
\bibitem[Wang and Dey(2010)]{WD10} Wang, X. and Dey, D. K. (2010). Generalized extreme value regression for binary response data: an application to B2B electronic payments system adoption. {\sl Annals of Applied Statistics} 4: 2000--2023.
\end{thebibliography}
\end{document}